# Highly ordered LIPSS on Au thin film for plasmonic sensing fabricated by double femtosecond pulses

Fotis Fraggelakis[1,*], Panagiotis Lingos[1] , George D. Tsibidis[1] , Emma Cusworth[2] , N. Kay[2], L. Fumagalli[2], Vasyl G. Kravets[2] , Alexander N. Grigorenko[2] , Andrei V. Kabashin[3] and Emmanuel Stratakis[1,4]*

[1]Institute of Electronic Structure and Laser (IESL), Foundation for Research and Technology (FORTH), N. Plastira 100, Vassilika Vouton, 70013 Heraklion, Crete, Greece

[2]Department of Physics and Astronomy, Manchester University, Manchester M13 9PL, UK

[3]Aix Marseille Univ, CNRS, LP3, Campus de Luminy, Case 917, 13288, Marseille, France

[4]Department of Physics, University of Crete, 71003 Heraklion, Crete, Greece

*Correspondence: Fotis Fraggelakis: +302811391940, fraggelakis@iesl.forth.gr, Emmanuel Stratakis: +302810391274, stratak@iesl.forth.gr

## Abstract

Periodic plasmonic arrays making possible excitations of surface lattice resonances (SLRs) or quasi-resonant features are of great importance for biosensing and other applications. Fabrication of such arrays over a large area is typically very costly and time-consuming when performed using conventional electron beam lithography and other methods, which reduce application prospects. Here, we propose a technique of double femtosecond pulse (~ 170 fs) laser-assisted structuring of thin (~ 32 nm) Au films deposited on a glass substrate and report a single-step fabrication of homogeneous and highly ordered Au-based Laser Induced Periodic Surface Structures (LIPSS) over a large area. Our experimental results unveil the key importance of the interpulse delay as the determining factor rendering possible the homogeneity of laser induced structures and confirm that highly ordered, functional LIPSS occur solely upon double pulse irradiation under a specific interpulse delay range. A theoretical investigation complements experimental results providing remarkable insights on the structure formation mechanism. Ellipsometric measurements show that such LIPSS structures can exhibit highly valuable plasmonic features in light reflection. In particular, we observed ultranarrow resonances associated with diffraction-coupled SLRs, which are of paramount importance for biosensing and other applications. The presented data suggest that femtosecond double pulse structuring of thin metal films can serve as a valuable and low-cost tool for a large-scale fabrication of highly ordered functional elements and structures.

## I. Introduction

Relying on the control of biological binding events between a target analyte from a solution and its corresponding surface-immobilized receptor via refractive index (RI) monitoring, plasmonic biosensors present a leading label-free technology used for the detection of a variety of critically important analytes in tasks of biomedical diagnose, food and environmental monitoring, safety and security[1]. Plasmonic biosensors employ resonant optical excitation of free electron oscillations (plasmons) in noble metal nanostructures (typically, Au) and profit from their strong dependence on RI of the ambient liquid medium, which renders possible a real-time control of bioreactions and the determination of kinetic constants within minutes (such control is hardly possible using conventional label-based technology). However, despite the development of a series of promising designs based on Surface Plasmon Resonance (SPR)[2]and Localized Plasmon Resonance (LPR)[3], currently available plasmonic label-free biosensors still need an improvement of sensitivity to compete with labelling approaches. As we showed in our earlier studies, one of the most promising ways to improve the sensitivity of sensing transducers is related to the employment of novel artificial materials (metamaterials), composed of nanoscale blocks arranged in a lattice, which could outperform natural materials in terms of response to RI variations due to the involvement of novel physical phenomena (e.g., electromagnetic, near-field, far-field coupling)[4–6]. One class of such metamaterials is based on periodic nanodot/nanostripe arrays, which empower the excitation of ultranarrow plasmonic SLRs with the resonance width down to 1-2 nm[7,8]. Such ultranarrow SLRs have very high resonant quality factor $Q = \lambda_{min}/\Delta\lambda > 200$[8] and thus enable high sensitivity and much higher precision of measurements in spectral interrogation compared to conventional SPR and LPR by profiting from the well-resolved ultranarrow SLR feature. In addition, when the light intensity at the SLR minimum is nearly zero (light darkness), one can observe singular (Heaviside-like) behavior of phase of reflected light, which opens access to extreme biosensing sensitivity if phase is employed as the sensing parameter [5,6,9]. Such metamaterial-based nano-transducers give a great promise for a radical upgrade of current biosensing technology, but their active exploitation is complicated by necessity of using expensive electron beam lithography (EBL) and focused ion beam (FIB) fabrication techniques, which are hardly compatible with the desired low-cost and scalability of sensing devices.

We believe that the bottleneck problems in the fabrication of large-scale functional plasmonic arrays can be overcome by the elaboration of a laser processing technique, which is fast, single step, inexpensive and easily scalable. Such a technique implies a multi-pulse irradiation of a target by ultrashort (mostly femtosecond) focused laser beam, which leads to a spontaneous formation of various nanoarchitectures on the target surface. This technique has already been widely used to nanostructure bulk targets to report the formation of a variety of micro/nano morphologies with features covering a broad palette of shapes, sizes and hierarchical formation[10]. Depending on the conditions of laser processing and the type of the material, such morphologies can be tailored to enable a variety of novel functionalities, including coloring[11], superhydrophobicity[12], anti-icing[13], anti-reflectivity[14], surface blackening[15] altered bacteria[16] and cell adhesion[17]. Laser-induced periodic surface structures (LIPSS) present the most common type of laser-processed structures that can be reproduced for almost any kind of material. Particularly, the Low Spatial Frequency LIPSS (LSFL)[18] are periodic structures with sizes comparable to the laser wavelength and can have either uniaxial (1D-LIPSS) or multiaxial symmetry (2D-LIPSS). On the contrary, LIPSS with periodicities well below the laser wavelength ($\lambda/2$ to $\lambda/10$) are coined as high spatial frequency LIPSS (HSFL)[18]. The formation of all these architectures can be achieved via the variation of laser parameters (wavelength, polarization, fluence and number of incident pulses, etc.). In particular, the LSFL period is linked to the laser wavelength, whereas the HSFL period seem to mostly depend on other parameters, such as the laser fluence and the pulse duration[19]. At the same time, the final structure formation is a multi-pulse process, in which the surface relief is shaped progressively pulse after pulse[20], therefore a certain number of pulses is required for developing pronounced structures. When large areas are textured, the delivery of evenly distributed amounts of energy over the processed area is essential to maintain the homogeneity of the structures formed, which makes the irradiation strategy a crucial aspect of the process. More complex morphologies arise when pulse characteristics such as polarization, spatial intensity and temporal profile are tailored. For example, 2D-LIPSS can be fabricated upon irradiation by beams with circular[21] and azimuthal/radial polarization[22].

We hypothesize that under some conditions LIPSS could provide promising nanoarchitectures for the implementation of ultrasensitive plasmonic transducers. However, biosensing applications of such transducers would require the generation of LIPSS on thin metal films (not on bulk substrates), which presents a novel challenging task. Compared to the bulk material case, major differences are anticipated for laser structuring of thin films due to a significantly different ablation geometry and related physical phenomena[23]. First, in the case of thin films the material's thickness is comparable with the penetration depth of the laser radiation, which results in a modification of the energy absorption profile[23]. Second, a processing geometry, in which an irradiated thin metal film is sandwiched between two dielectric media (air and glass in our case), can lead to the excitation of electromagnetically coupled Surface Plasmon Polaritons (SPPs) over two opposite sides of the metal film, resulting in a complex spatial modulation and re-distribution of the laser energy within the material[24–26]. Third, laser structuring of thin metal films is often accompanied by a delamination, substrate influence on thin film adhesion and thermal stresses, which can lead to spallation and removal of irradiated material. To overcome these problems during LIPSS formation, one typically has to carefully control pulse energy and the total number of pulses[23]. Finally, the development of plasmonic devices is mostly based on noble metals, in which LIPSS are difficult to form upon single pulse irradiation[27]. Indeed, especially for thin Au films, the period of the structures as well as their homogeneity is difficult to control while delamination at the edge of structures is evident[28].

We believe that the above-stated limitations can be overcome by tailoring the optical energy deposition process via a temporal beam shaping similarly to how it was previously reported in experiments on double pulse structuring of bulk dielectric or semiconductor targets, which made possible an efficient control of LIPPS formation[29,30]. Here, we explore the use of double pulse femtosecond laser structuring of a glass substrate-supported thin (32 nm) Au film in both picosecond (ps) and nanosecond (ns) interpulse delay regimes and study the impact of important process parameters such as laser fluence, pulse to pulse overlap and pulse delay ($\Delta\tau$) on properties and quality of the formed structures. We show that under some conditions such a structuring can indeed lead the fabrication of homogeneous high-quality LIPSS on thin Au films, which render possible the generation of prominent plasmonic phenomena. A theoretical model is also employed to interpret the impact of the underlying physical processes in the topography formation. Ellipsometric measurements are conducted to evaluate whether the fabricated LIPSS have sufficient quality to provide ultranarrow SLRs with related light phase jump features at resonance minima, which can be used in ultrasensitive plasmonic biosensing. The detailed investigation confirms our original hypothesis on the possibility of fabrication of low-cost and scalable optical nanotransducers using the technique of laser processing.

## II. Results and discussion

A detailed investigation of the topographies obtained following single and double pulse laser irradiation of Au thin films is presented and discussed. The laser processing results are complemented with theoretical simulations on laser surface coupling and surface functionality characterization.

### *a) Resulting Morphology for Single and Double Pulse Structuring*

Figure 1 illustrates the resulting morphology on Au surface after irradiation with single and double pulses, respectively. LIPSS formation is a multi-pulse process where different combinations of the laser pulse overlap (*Ov*) and peak fluence $\Phi$ lead to different morphologies. Figure 1 shows the induced morphology for variable $\Phi$ and pulse separation ($\Delta\tau$) values when the pulse-to-pulse overlap is fixed to *Ov* = 150 pulses per spot (pps). Different fluence ranges are considered for single and double pulses. For single pulses, the fluence values vary between $\Phi$ = 140 and 200 mJ/cm$^2$, while for double pulses between $\Phi$ = 160 and 240 mJ/cm$^2$. It is worth emphasizing that $\Phi$ refers to the total (peak) fluence of the pulse set in case of double pulses. The fluence values are selected appropriately to illustrate the different laser induced morphologies from structure appearance (low $\Phi$) to thin film damage (high $\Phi$). The difference in the $\Phi$ values between single and double pulses is attributed to the variation of the effective fluence between single- and double-pulse structuring as well as between different $\Delta\tau$ values[31]. The source repetition rate is 5 kHz corresponding to interpulse time $T$~ 200 µs, several orders of magnitude larger than surface resolidification time upon fs irradiation[32]. Therefore, its influence on the ripple formation process is neglected.

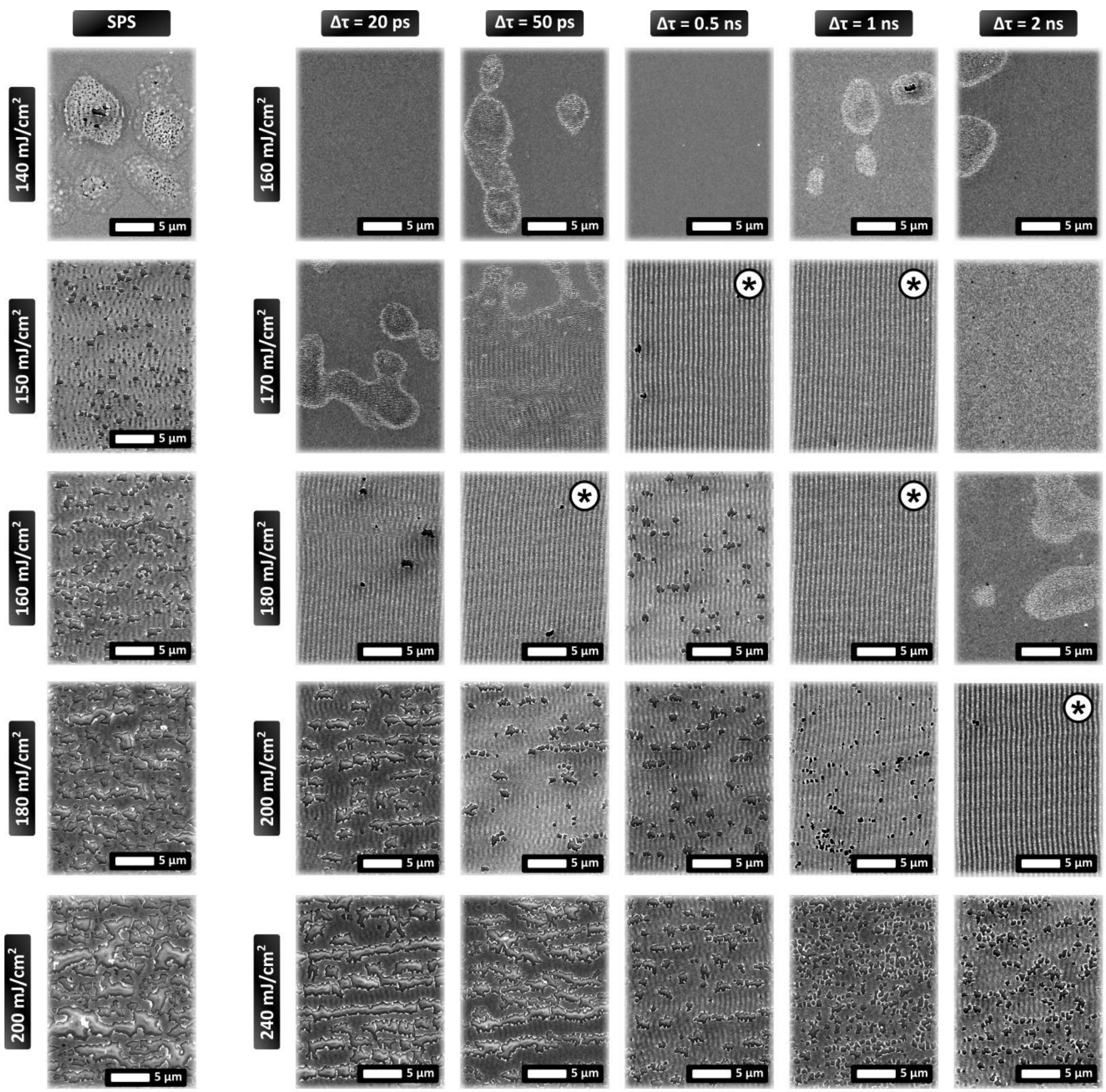

Figure 1: i) SEM images showing the different morphologies obtained by SPS and DPS structuring at different fluences, *Φ*, and interpulse delays, *Δτ*. Homogeneous LIPSS marked by '*'. An overlap *Ov*=150 pulses per spot (pps) is used.

**Single Pulse Structuring – SPS**

For single pulse structuring (Figure 1, SPS and $\Phi$ = 140 mJ/cm$^2$) the material is only partially structured; more specifically, small size craters are formed on the surface, while traces of LIPSS form around the crater. For $\Phi$ = 150 mJ/cm$^2$ craters densify in a random way on the surface, their size is of the order of 1 µm while the areas between them is covered with inhomogeneous and pale LIPSS. For $\Phi$ = 160 mJ/cm$^2$ craters densify further and merge, while LIPSS appear to be less prominent compared to $\Phi$ = 150 mJ/cm$^2$. For higher fluences ($\Phi$ = 180 - 200 mJ/cm$^2$) the craters grow in number and size leading to film damage.

**Double pulse structuring – DPS**

For double pulse structuring (DPS) and $\Delta\tau$ = 20 ps, a different evolution of structures is observed upon increasing the fluence. For $\Phi$ = 160 mJ/cm$^2$ no trace of surface modification is observed, while for $\Phi$ = 170 mJ/cm$^2$ areas with nano-sized roughness appear on the surface. Notably, for $\Phi$ = 180 mJ/cm$^2$, prominent and homogeneous LIPSS are formed together with few craters (Fig. 1). Upon an increase of fluence to $\Phi$ = 200 and 240 mJ/cm$^2$, craters densify, grow and merge in a similar way as for SPS leading to a surface damage. It is essential to note here that the beating effect reported on thin Au structuring[28] is not observed in our experiments either in SPS or in DPS case, the ripple period has only two distinct values. This could be attributed to different effective fluences that lead to area film removal, as well as to different thicknesses of material (6-20 nm in Takami et. al[28] and 32 nm here).

A similar trend is observed in all the cases of DPS presented in Figure 1; more specifically, at low $\Phi$ values, random areas with nano roughness appear on the surface. At intermediate $\Phi$ values, homogeneous LIPSS are formed on the surface for most of the interpulse delays examined. As $\Phi$ increases further, craters are formed, densify, grow and damage the surface. The specific fluence threshold leading to the formation of the morphologies varies depending on the interpulse delay. Homogeneous and prominent LIPSS, marked by '*' in Figure 1, are formed almost in all cases of DPS. $\Phi$ values that lead to homogeneous LIPSS will be referred as *Best Fluence* ($\Phi_{best}$). $\Phi_{best}$ varies for each $\Delta\tau$. In particular, for interpulse delays of 50 ps, 0.5 ns, 1 ns and 2 ns, LIPSS are formed at $\Phi_{best}$ values of 180 mJ/cm$^2$, 170 mJ/cm$^2$, 170-180 mJ/cm$^2$ and 200 mJ/cm$^2$, respectively. The observed variation of the $\Phi_{best}$ values leading to optimum LIPSS structures is attributed to both the SPP excitation and the ultrafast dynamics affected by the interpulse delay. The contribution of each of these two mechanisms will be elucidated by the experimental and theoretical data provided in the following sections.

### b) $\Delta\tau$ Impact on H-LIPSS Process Window

Since the morphologies studied here are the result of multi-pulse irradiation, **both *Φ* and *Ov* have an impact on the induced morphology**. Therefore, the contribution of *Φ* and *Ov* should be analyzed separately bringing in a new perspective in the process window investigation. Some combinations of *Φ* and *Ov* result in homogeneous LIPSS formation (termed here as *H-LIPSS)*. Other possible combinations lead to inhomogeneous LIPSS generation, coined as *partially formed LIPSS* formation or the formation of *LIPSS & craters*. Combinations of relatively low values of *Φ* and *Ov*, do not suffice to melt the material and therefore, the surface remains *untextured*; Solely when $\Delta\tau$ = 2 ns, there are some combinations leading to the fabrication of *Dual period LIPSS* where the induced patterns are characterized by two different periods. The above morphologies for different delays are presented using color-coding in the process window graphs of Figure 2. More specifically, the combination of the process parameters that lead to *untextured* surface is indicated with non-filled circles. Textured surfaces with *partially-formed LIPSS, H-LIPSS* [1], *LIPSS & craters,* and *Dual Period LIPSS* are represented in Figure 2 with circles filled with a *light yellow, yellow, brown,* and *pink* color, respectively. Finally, delaminated or crater-dominated areas are marked with circles filled with a *grey* color. In particularly, the morphology of the *Dual Period LIPSS* will be discussed in detail in the following section (*c)*.

A thorough investigation of the graphs in Figure 2, reveals that for a given $\Delta\tau$, $\Phi_{best}$ variance is dependent on *Ov*. H-LIPSS are formed for relatively low fluence values combined with relatively high pulse overlap, average fluence values combine better for average pulse overlap, as well as high fluence values with low pulse overlap. Moreover, considering the yellow shaded areas, the transition from non-textured surface to H-LIPSS formation ultimately to surface with LIPSS & craters in a non-abrupt way as the dose increases. Yet, the average dose, *D*, is substantially different in each case and increases upon fluence decrease. For example, when $\Delta\tau$ = 1.5 ns H-LIPSS are formed for *D* = 7.2 J/cm$^2$ when $\Phi_{best}$ = 240 mJ/cm$^2$ (*Ov* = 30 pps) and *D*= 68 J/cm$^2$ when $\Phi_{best}$ = 170 mJ/cm$^2$. *D* value varies in this case approximately one order of magnitude pointing out that H-LIPSS formation occurs above a certain *Φ* threshold. When *Φ* is low (*Φ* = 130 mJ/cm$^2$), that threshold is reached at a smaller spot radius and many pulses is required to texture the area (*Ov* = 400 pps). When the fluence is high (*Φ* = 200 mJ/cm$^2$), the effective spot radius is larger, and a small number of pulses is required to texture the area (*Ov* = 30 pps). Thus, relatively higher *Φ* values are more effective for LIPSS formation and this is the trend observed for any $\Delta\tau$ value tested.

A comparison of the relevant process windows for different interpulse delays demonstrates the role of $\Delta\tau$ as an important optimization parameter. It is shown that apart from the *Ov*&*Φ, Δτ* value appears to have a remarkable effect on the H-LIPSS process window size. The number of combinations of *Ov*&*Φ* leading to H-LIPSS formation differs significantly with respect to the interpulse delay ($\Delta\tau$). In particularly, in the process window graphs presented in Figure 2, the number of different *Ov*&*Φ* combinations leading to H-LIPSS formation are illustrated for each $\Delta\tau$ . It is shown that H-LIPSS occurrences increase from four to ten (see legend in each graph) as the delay increases from $\Delta\tau$ = 0.05 ns to 1 ns before a drop to six occurrences are observed . The incidents remain 10 for $\Delta\tau$ = 1 ns and $\Delta\tau$ = 1.5 ns and then reduce to 6 for $\Delta\tau$ = 2 ns.

[1] Presence of less than 5 craters with diameter larger than 500 nm within an area of 500 μm$^2$

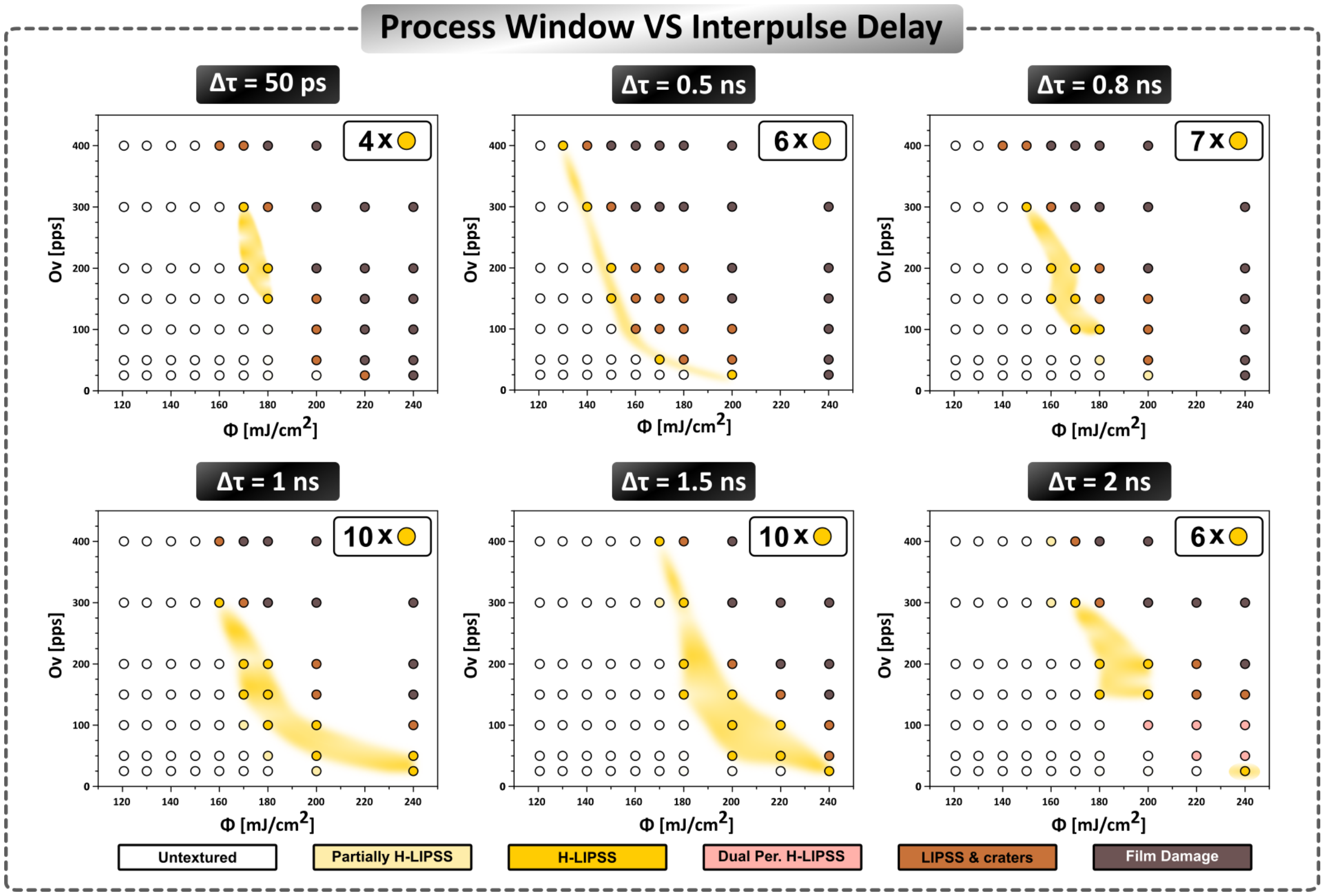

Figure 2. Process window graphs indicating the resulting morphology upon systematic variation of two process parameters (*Φ & Ov*) for six interpulse delay (*Δτ*) values. The colour of the circle fillings corresponds to the different morphology attained for each set of parameters as indicated in the legend shown at the bottom. The yellow marks in each graph indicate the process parameters that lead to homogeneous (H)-LIPSS formation. The shaded yellow areas in between the graphs are the estimated process windows for different delays. The number of cases of H-LIPSS formation observed within the process window are indicated at the top right of each graph.

### *c) SPS VS DPS: Structure Evolution and Optimized Morphologies*

**Structure Evolution**

In the previous sections (II.a and II.b), remarkable differences between SPS and DPS efficiencies to generate H-LIPSS on thin Au films were presented for distinct process parameters. Next, the features of the surface topography as a function of *Ov* will be discussed to elucidate the differences in the surface pattern features for SPS and DPS. Starting from originally unprocessed surface (Figure 3i), several defects or imperfections on the order of 300 nm in size have been observed. Upon applying SPS with $\Phi$ = 140 mJ/cm$^2$ and *Ov* = 50 pps (Figure 3ii, SPS), small-sized (i.e. hole-like) craters are produced on the surface together with a few obscure periodic structures surrounding them and some defect features (turquoise arrow in Figure 3ii, SPS, 50 pps). Upon increasing the energy dose (*Ov* = 100 pps), concentric periodic features similar to LIPSS are formed around the craters while the presence of the defects starts to diminish. The similarity between the density of the defects on the unprocessed surface and the induced craters in SPS indicate that the imperfections act as hot spots which result in a local amplification of the intensity of the incident laser field[33]. Following the increase of the overlap, (*Ov* = 150 pps) the number of craters increases significantly and LIPSS surrounding them become more pronounced. Under these conditions, LIPSS of adjacent craters overlap. Since the craters are distributed randomly following the substrate's imperfections, their distance is not always an integer multiple of the SP wavelength. Therefore, when the textured areas overlap, regions covered with quasi-periodic structures are produced (ellipse in *green* color in Figure 3ii, SPS, *Ov* = 150 pps). The appearance of craters preceding regular LIPSS formation seems to prevent H-LIPSS generation for SPS.

In contrast, for the DPS, the surface evolution upon increase of *Ov* is significantly different compared to SPS. The pattern produced for DPS is illustrated in Figure 3ii for $\Delta\tau$ = 1 ns. The delay value lies in the center of the optimum $\Delta\tau$ range (0.8 ns<$\Delta\tau$< 1.5 ns) for H-LIPSS formation as discussed in previous paragraph. The corresponding SEM images shown in Figure 3ii, indicate that for an initial *Ov* = 100 pps more imperfections are generated on the irradiated surface which are of the same size with those in the unprocessed one. For *Ov* = 200 pps, the roughness of the surface increases followed by an increase of the number of the imperfections; notably, the size of the defects appears to grow slightly. Finally, at *Ov* = 300 pps, regular LIPSS are formed throughout the irradiated area, while no imperfections are observed. EDX measurements (see Supplementary Material) confirmed a nearly complete removal of the Au film between LIPSS in some of the examined areas for *Ov* = 300 pps (Figure 3ii). The Au film removal could be attributed to either a microfluidic movement or ablation related processes.

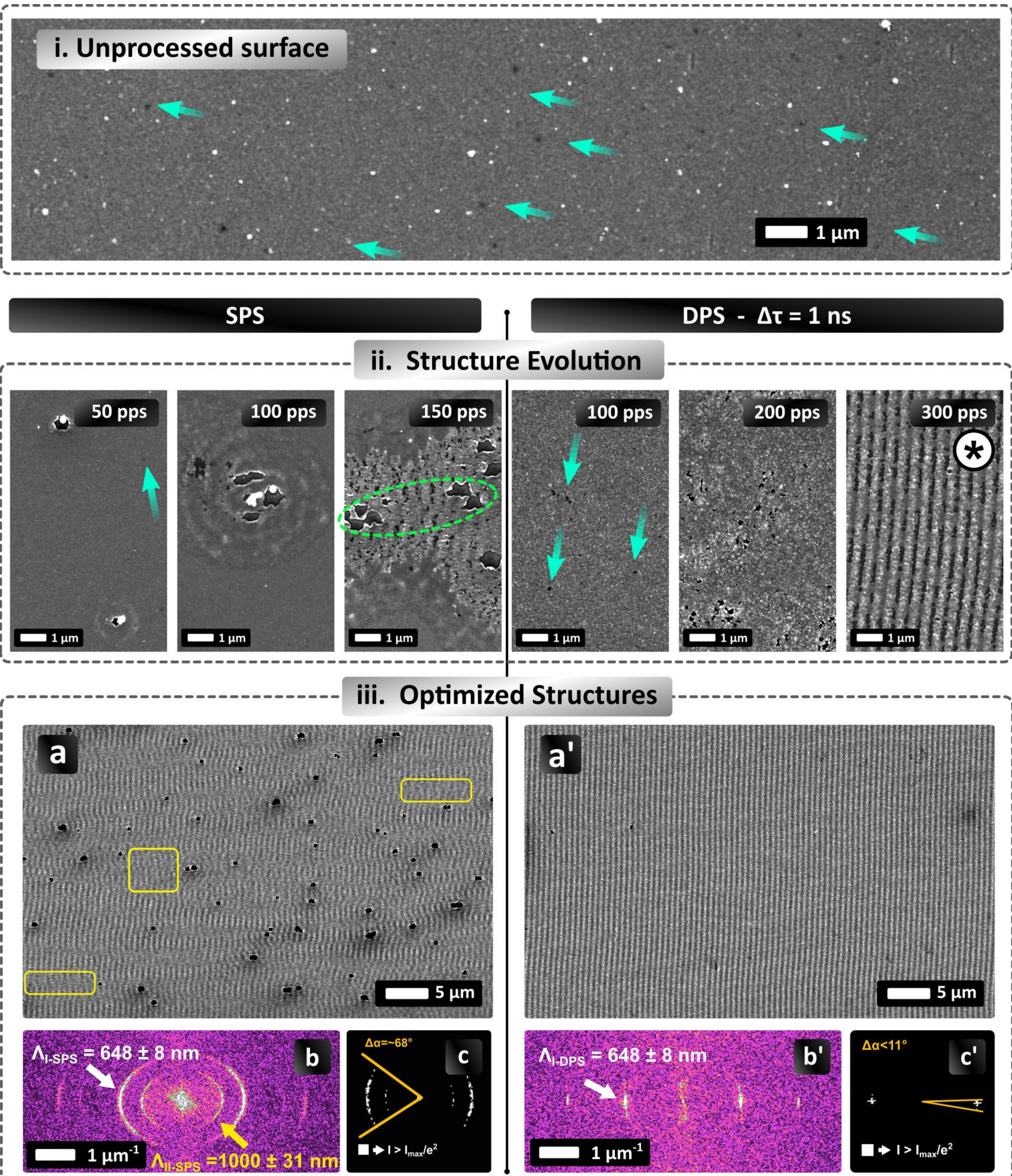

Figure 3 Comparison of structures for SPI and DPI ($\Delta\tau$ = 1 ns). i) Unprocessed surface. Turquoise arrows indicate surface defects. ii) Comparison of structure evolution for SPS and DPS respectively. *Ov* is indicated for each SEM image $\Phi$= 140 mJ/cm$^2$ for SPI and $\Phi$= 160 mJ/cm$^2$ for DPI. Turquoise arrows indicate the “hot spots” present on the surface and green dashed ellipse the inhomogeneous morphology between two craters. Symbol “*” indicates the surface subjected to EDX measurement provided in supplementary material. iii) a and a’: SEM of optimized surfaces for SPS and DPS. For SPI, $\Phi$ = 180 mJ/cm$^2$ and *Ov* = 100 pps whilst for DPI, $\Phi$ = 170 mJ/cm$^2$ and *Ov* = 150 pps. b and b’: Fast Fourier transform (FFT) maps are shown for SPI and DPI respectively. c & c’: FT maps showing the signal points with intensity $I>I_{max}/e^2$ of b and b’ respectively.

**Optimized Morphologies**

A key strategy for a controlled development of functional patterns for potential applications requires the capacity to fabricate optimized morphologies over a large area. The term 'optimized morphology' is related to the production of textured surfaces, which are fine-tuned and designed for peak performance according to the application. In the current study, the "best" or "optimized" topography is associated with a pattern characterized by an enhanced homogeneity (C1), spatial coherence of structures (C2), undispersed directionality (C3), and a single period (C4) over the processed area. For example, the morphology shown in Figure 3iii, fails to meet any of those conditions: the induced pattern is covered with structures that are incoherent and have multiple periods; in addition, the topography is inhomogeneous due to the presence of craters on the surface. Regarding the capacity to produce patterns satisfying the above requirements, a textured surface shown in Figure 1 for $\Delta\tau$ = 0.5 ns at $\Phi$ = 170 mJ/cm$^2$ meets conditions (C2-C4) but not condition (C1) due to the presence of small-sized craters.

In Figure 3iii, a and a', the 'best' possible morphology obtained employing SPS and DPS, respectively, is illustrated. A Fast Fourier Transform (FFT) analysis was employed (Figure 3iii, b and b') to determine the pattern periods. To evaluate the directionality (or "angular dispersion" of the structures), the 2D - FFT maps (Figure 3iii, b & b') were subjected to a threshold analysis. The threshold value was set to 1/e$^2$ of maximum signal intensity in FFT for the dominant period $\Lambda_I$~640 nm and a graph was produced in each case, showing in white the points exceeding this threshold (Figure 3iii, c and c').

The above analysis demonstrates that the generation of a surface satisfying all conditions (C1-C4) is not possible in the case of SPS. More specifically, one of the characteristic examples of the SPS-obtained morphology is shown in Figure 3iii, a where several defects are formed on the surface coexisting with LIPSS, which cover partially the irradiated area (C1 is not met). LIPSS are formed around the craters and therefore, there is a spatial incoherence (C2 is not met). Spatial incoherence is more evident in regions between structured areas, where ripples interfere (yellow boxes in Figure 3iii, a). An FFT analysis (Figure 3iii, b) shows the formation of LIPSS of two periods ( $\Lambda_{I\text{-}SPS}$ = 648 ± 8 nm and $\Lambda_{II\text{-}SPS}$ = 1000 ± 31 nm) while condition (C4) is not satisfied. Furthermore, an analysis of the signal intensity provides an estimation of the angular dispersion of the structures, namely, Δα ~68° (Figure 3iii, c) .

In contrast, for DPS at $\Delta\tau$ = 1 ns, defect-free (C1) and spatially coherent (C2) LIPSS (H-LIPSS) areas are formed (Figure 3iii, a'). An FFT analysis for DPS, $\Delta\tau$ = 1 ns (Figure 3iii, b'), indicate a single period of $\Lambda_I$ = 648 ± 8 nm. In both cases, the predominant LIPSS period is approximately equal to 650 nm. The angular dispersion in the case of DPS is estimated to be equal to Δα ~11° (Figure 3iii, c'). Therefore, patterns obtained via DP, satisfy all the conditions of 'optimised' morphologies in contrast to structures produced via SPS.

### *d) Discussion on Structure Formation Mechanism*

**Formation Mechanism Overview**

The observed remarkable differences between the patterns induced via single and double pulses stimulate the clarification of a set of physical phenomena that accompany laser-matter interaction:

Specifically the (i) formation of small imperfections/defects on the topography which are more pronounced for single pulses (or for double pulses of high fluence), (ii) explanation of the occurrence of two LSFL periodicities on the surface of the irradiated solid, (iii) establishment of the influence of long (ns) delays and the interplay of fluence and pulse separation on the characteristics of the produced patterns, and (iv) clarification of the role of the film thickness on the size of the observed periodicities. To address the above challenges, a consistent strategy for the description of the physical mechanisms that lead to surface patterning in various laser conditions is required; such an approach should include a detailed investigation of multiscale (temporal) phenomena including energy absorption, electron excitation and relaxation processes, phase transition and pattern formation. The emphasis on the present work is on the interpretation of the formation of LIPSS for single and double pulses.

In principle, the formation of laser-induced periodic structures in metallic surfaces is mainly attributed to two complementary mechanisms: (i) The excitation of Surface Plasmon Polaritons (SPP) on the metal-dielectric interface resulting from scattering of the incident light on a rough surface[26,34,35], which leads to an inhomogeneous laser intensity absorption (Process I), (ii) a hydrodynamic movement of a molten material driven by temperature gradients and dictated by out-of-equilibrium flow [36], which is also a sufficient condition for the onset of surface waves that will lead eventually to periodic patterns [20,37–39](Process II). Different timescales of the two processes have been unveiled by both pump-probe experiments[40,41] and simulations[42] in previous reports. In the following sections we will present experimental and theoretical data to describe the distinct role of each of the two processes in the structure formation. A special emphasis will be placed on the formation of the Dual Period LIPSS.

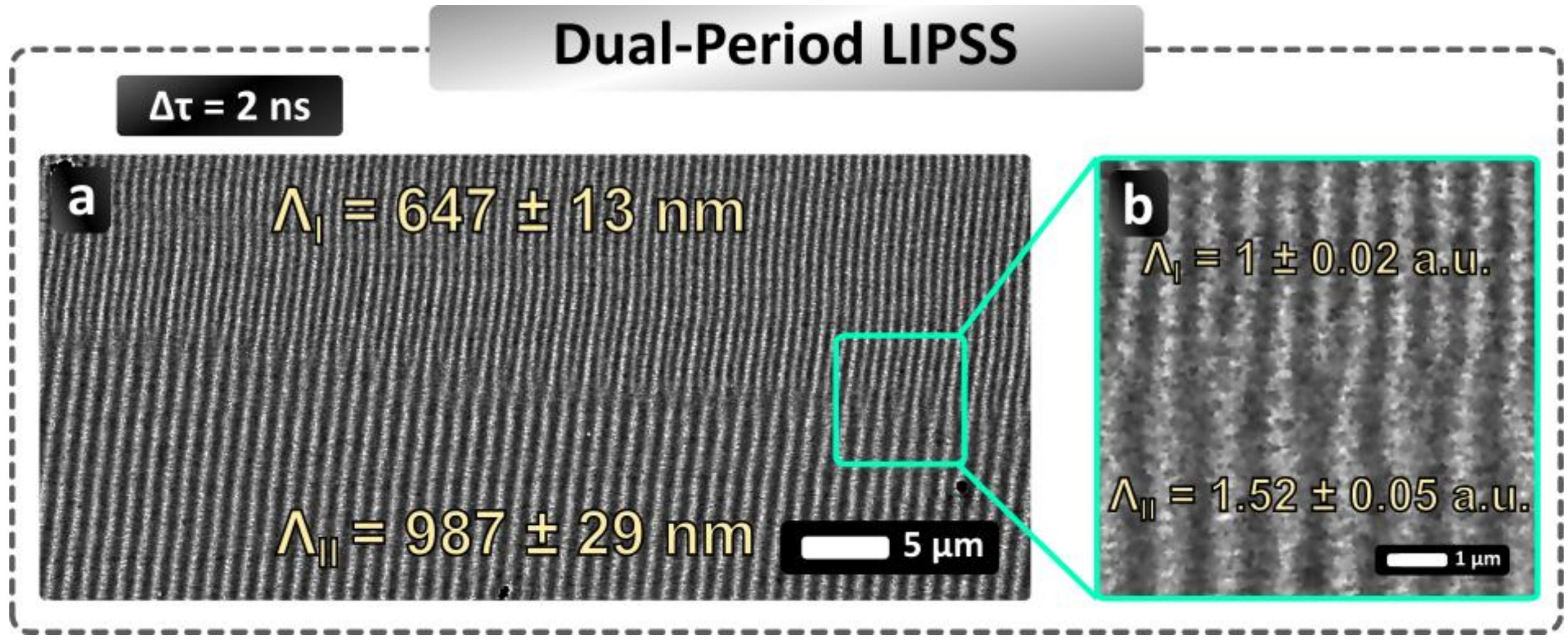

Figure 4: a,b) SEM image of Au surface showing the particular formation of dual-period LIPSS after irradiation by double pulses having $\Delta\tau$ = 2 ns, N = 100 pps and Φ = 200 μJ/cm$^2$ . A direct comparison figure showing Ref[36] *Plate 7 can be found in the published version of the manuscript.*

**Dual-Period H-LIPSS**

A remarkable experimental observation for DPS is the production of patterns characterized by the coexistence of homogeneous (C1), spatially coherent (C2) and unidirectional (C3) LIPSS that have two different periods in adjacent areas (not C4). An example of such a topography is illustrated in Figure 4,a and b. Based on the analysis of the experimental data, Dual Period LIPSS were observed *only* for $\Delta\tau$ = 2 ns (see circles filled in pink color in Figure 2). The measured two LIPSS periods have two distinct values $\Lambda_I$ = 647 ± 13 nm and $\Lambda_{II}$ = 987 ± 29 nm (Figure 4a) which are comparable to the values of the LIPSS observed for SPS, $\Lambda_{I\text{-}SPS}$ = 648 ± 8 nm and $\Lambda_{II\text{-}SPS}$ = 1000 ± 31 nm (Figure 3ii, b); the lower value in these two cases is also approximately equal to the period observed for $\Delta\tau$ = 1 ns, namely $\Lambda_{I\text{-}1ns}$ = 648 ± 8 nm (Figure 3ii, b'). On the other hand, $\Lambda_I$ values are similar to the theoretically calculated values of SPPs induced in metal/glass ($\Lambda_{bottom}$ = 690 nm) and $\Lambda_{II}$ for the SPPs induced in air/metal ($\Lambda_{top}$ = 1026 nm) interfaces (see Supplementary material) that indicates a potential contribution of the SPP in the formation of the Dual Period LIPSS.

Although Dual Period LIPSS are observed upon both the employment of SPS and DPS, their homogeneity and directionality differ remarkably between the two cases. For SPS, these LIPSS are observed sporadically and are overlapped. For DPS and specifically when $\Delta\tau$ = 2 ns, Dual Period LIPSS are formed over extended areas with a spatial separation of regions where LIPSS of lower and larger periodicity are formed (Figure 4,a). Notably, several combinations of $\Phi$ and *Ov* are capable to lead to Dual Period LIPSS, ranging between *Ov* = 50 – 100 pps and $\Phi$ = *200 – 240* mJ/cm$^2$. Thus, a low overlap of pulses and high fluences within the process window are required to produce Dual Period LIPSS).

**The Convection Flow hypothesis**

Even though the origin of the observed periodicities can be described quite accurately by considering the excitation of SPPs on the two interfaces (Air/metal an Metal/Substrate), the underlining mechanism leading to the coexistence of the two periods in the homogeneous Dual Period LIPSS solely upon DPS appears to be elusive. A potential interpretation of the relevant process should be attributed to the role of hydrodynamical phenomena (Process II). In Figure 4, a region of the processed surface is illustrated to demonstrate a transition area between the two different LIPSS periods. The obtained LIPSS pattern resembles the profile of a particular type of convection flow identified as "Pinching Instability" shown in Ref[36] *Plate 7 - Figure 14*. In the Pinching Instability, two distinct periods of convection flow rolls are produced to form a homogeneous, unidirectional, spatially coherent and spatially separated structure. Interestingly, the ratio between the small and the large period is similar both in Dual Period LIPSS and in the Pinching Instability (between Figure 4 and Ref[36] *Plate 7*). A comparison of the ratios can be made by normalizing the ratios in the two cases. The nominal value (1 a.u.) is the smaller value of each case ($\Lambda_I$ and $\Lambda_I'$). The two normalized LIPSS periods on laser processed Au (Figure 4b) are $\Lambda_I$ = 1 ± 0.02 a.u and $\Lambda_{II}$ = 1.52 ± 0.05 a.u. corresponding to $\Lambda_I$ = 647 ± 13 nm and $\Lambda_{II}$ = 987 ± 29 nm in Figure 4a. The two normalized periods for the Pinching Instability, measured in Ref[36] *Plate 7* are $\Lambda'_I$ = 1 ± 0.07 a.u and $\Lambda'_{II}$ = 1.41 ± 0.11 a.u. Those values do overlap within the limits of the error pointing out that convection flow could potentially play a role together with the SPP excitation in the formation of dual-period LIPSS over large areas. Significant similarities between the obtained structures and convection flow patterns have also been reported on steel upon DPS[37], while theoretical works[20,43] describe an important contribution of Marangoni instability and convection flow to the LIPSS formation. According to such theoretical works, the instability pattern strongly depends on the local excitation conditions. In this context, we hypothesize that a periodic heat distribution on the melted material can impose a double period, which can hydrodynamically coexist only upon DPS due to convection flow. Moreover, for $\Delta\tau$ values close or above 2 ns, the resolidification time is abruptly reduced from *t*>30ns to *t*< *Δτ* and that would expect to have severe impact to Process II evolution. A distinct characteristic of convection flow is that the flow pattern dictating the material displacement is changing in time [36] . *Considering a constant evolution of the flow pattern and a sharp reduction in resolidification time we can argue that the double period LIPSS formation is a result of an early stage resolidification. In the same hypothesis, H-LIPSS formation can be interpreted as follows: H-LIPSS is the result of a fully developed, long-lasting convection flow that passes from a "pinching instability" dual-period pattern (t~2 ns), and it is later on stabilized to a single period homogeneous flow (t~30ns).*

### *e) Multi-physical Simulation Results*

In order to further evaluate the contributions of the electromagnetic (Process I) *vs* hydrodynamic (Process II) mechanisms in the formation of the observed structures, simulations were performed assuming the employment of a 1026 nm linearly polarized pulsed laser light with pulse-to-pulse duration of 170 fs irradiating a 32 nm thick Au film placed on a $SiO_2$ substrate. A multiscale analysis of the underlying physical phenomena for a single pulse of (peak) fluence *Φ*= 280 mJ/cm$^2$ and a double pulse (considering two constituent pulses of equal peak fluence, 140 mJ/cm$^2$, separated by 2 ns) was conducted. The fluence value was appropriately selected to ensure that the second pulse 'sees' a sufficiently large volume of molten Au and therefore, the interaction of a laser pulse with a molten material rather than a solid is examined.

Firstly, simulations are aimed to model the formation of Dual Period LIPSS. To explore the features of the excited surface waves (which are the precursors of both the periodicity value and the orientation of LIPSS) on an initially non-flat surface, the roughness on Au is emulated with a random distribution of a number of semispherical bumps of radius *R*=16 nm (for a more detailed description of the method see Ref.[26]; other relevant work is presented in Ref.[34]). Based on the electromagnetic simulations, the origin of the two observed LIPSS periods are attributed to the SPP periods, excited at the Au-substrate and air-Au interfaces, respectively. More specifically, far- and near- field excited modes are illustrated in Figure 5a (the figure indicates the intensity enhancement as a result of scattering of the incident beam off the bumps). An FFT analysis of Figure 5a is shown in Figure 5b in the $k_x/k_0(\approx \lambda/\Lambda)$ space ($\lambda, \Lambda$ correspond to the laser wavelength and the periodicity of the modes, respectively) which yields two periodicities attributed to the SPP excited on the two interfaces ($\Lambda_{bottom} = 690\ nm, \Lambda_{top} = 1026\ nm$ ). Interestingly, both the orientation (perpendicular to the laser polarization) and the periodicity values agree with the experimental observations (i.e. LSFL periods equal to $\Lambda_I$ = 647 ± 13 nm and $\Lambda_{II}$ = 987 ± 29 nm). The two calculated periodicities result from the interference of the incident beam with TM-polarized SPPs and quasi-cylindrical waves on air/metal interface and metal/dielectric interface. Similar predictions have been presented in a recent work [26] emphasizing the impact of the interplay of the two SPP which was also observed in a previous experimental study [28].

Nevertheless, the electromagnetic fingerprint itself does not suffice to explain the topography variation for single and double pulses. Thus, the intensity profile illustrated in Figure 5a is used to derive, firstly, the local energy absorption from the Au/$SiO_2$ two-layered material for single and double pulses.[26] The presence of the corrugation (see also Ref [26 ]) enhances locally the energy absorption which implies that not only the top SPP but also the bottom excited electromagnetic modes are expected to yield a pronounced effect projected on the air/metal surface. Furthermore, the electron excitation, thermal response of Au and topography formation (see more details in the Supplementary Material) are evaluated and remarkable changes are predicted between single and double pulses.

More specifically, it is shown that for a SPS (Figure 5c), the use of $\Phi$ = 280 mJ/cm$^2$ raises the lattice temperature to ablation levels (higher than the critical value of 5625 K) around the irregularity on the surface depicted by the region in *white* color. Thus, the analysis of the thermal effects of the irradiated Au leads, firstly, to a mass removal in the region covered with bumps while there is a translation of the electromagnetic fingerprint to the temperature profile. In particularly, the employment of a thermal model [see Ref. [26] and references therein] predicts that the entire Au surface undergoes a transition to a molten phase (Figure 5c shows the lattice temperature profile on the surface upper view at time *t*= 0.25 ns).

Moving to investigation of the physical mechanisms that dictate the pattern formation for a double pulse of total (peak) fluence $\Phi$=280 mJ/cm$^2$, the combined effect of the constituent pulses must be evaluated. In general, for DPS, the first pulse is expected to yield a similar electromagnetic profile to that of a single pulse (Figure 5a), while the maximum lattice temperature is evaluated to be well below the ablation point; this is due to the fact that the first constituent pulse (of peak fluence equal to 140 mJ/cm$^2$) is not capable to ablate the material. Figure 5d and e, illustrates the surface temperature referring to the case of, the maximum experimentally examined $\Delta\tau$ = 2 ns. The surface of Au will experience a phase transition as a result of the irradiation with the first pulse and therefore, the second pulse of the same energy will 'see' a molten material. It is also noted that the calculated maximum thickness of the molten volume is ~25 nm at the time the second pulse irradiates the material. The evaluation of the electromagnetic fingerprint requires an analysis of the electromagnetic response of the material to the irradiation of the second pulse. In comparison with the corrugation introduced on a solid surface, elongated islands of molten material of a spatially varying thickness represent the roughness (Figure 5d) at which the second pulse will scatter. Due to the resulting small temperature gradient (see Figure 5d) which depicts the lattice temperature profile at *t*=2 ns) the induced hydrothermal wave has a very small amplitude (~1 nm, see, also, Supplementary Material) which complicates significantly the calculation of the surface electromagnetic modes. Although a more precise evaluation is thus required, it is assumed that the temperature profile shown in (Figure 5e) can be used to approximately project the distribution of energy upon exposure of the $Au/SiO_2$ to the second pulse. Thus, the same combination of top and bottom SPP should be derived as for a single pulse or the first of the two pulses. Another aspect that has to be investigated is whether the SPPs excited from the first pulse contribute to the electromagnetic picture. Assuming the short SPP lifetime (of the size of some ps for noble materials[44]), the SPPs excited from the first pulse are not expected to interfere with the electromagnetic modes of the second pulse. Results shown at *t*~2.04 ns (Figure 5e) when the maximum temperature is attained illustrate the lattice temperature spatial profile. It is emphasized that due to the faster relaxation process for the double pulse, the height of the induced LIPSS is expected to lead to a more precise formation of LSFL structures (see Supplementary Material).

Applying the theoretical model for the case of $\Delta\tau$ = 1 ns, simulations predict the produced topography[45] for SPS and DPS (Figure 5f,g). It should be noted that in contrast to SPS (Figure 5f), the temperature never exceeds the ablation threshold and therefore for DPS and $\Delta\tau$ = 1 ns, there is no formation of ablation-assisted spots (dots in *orange* in Figure 5f) but only LIPSS (Figure 5g). These theoretical predictions are in agreement with the experimental observations discussed in the previous Section for $\Delta\tau$ = 1 ns (Figure 3ii and iii).

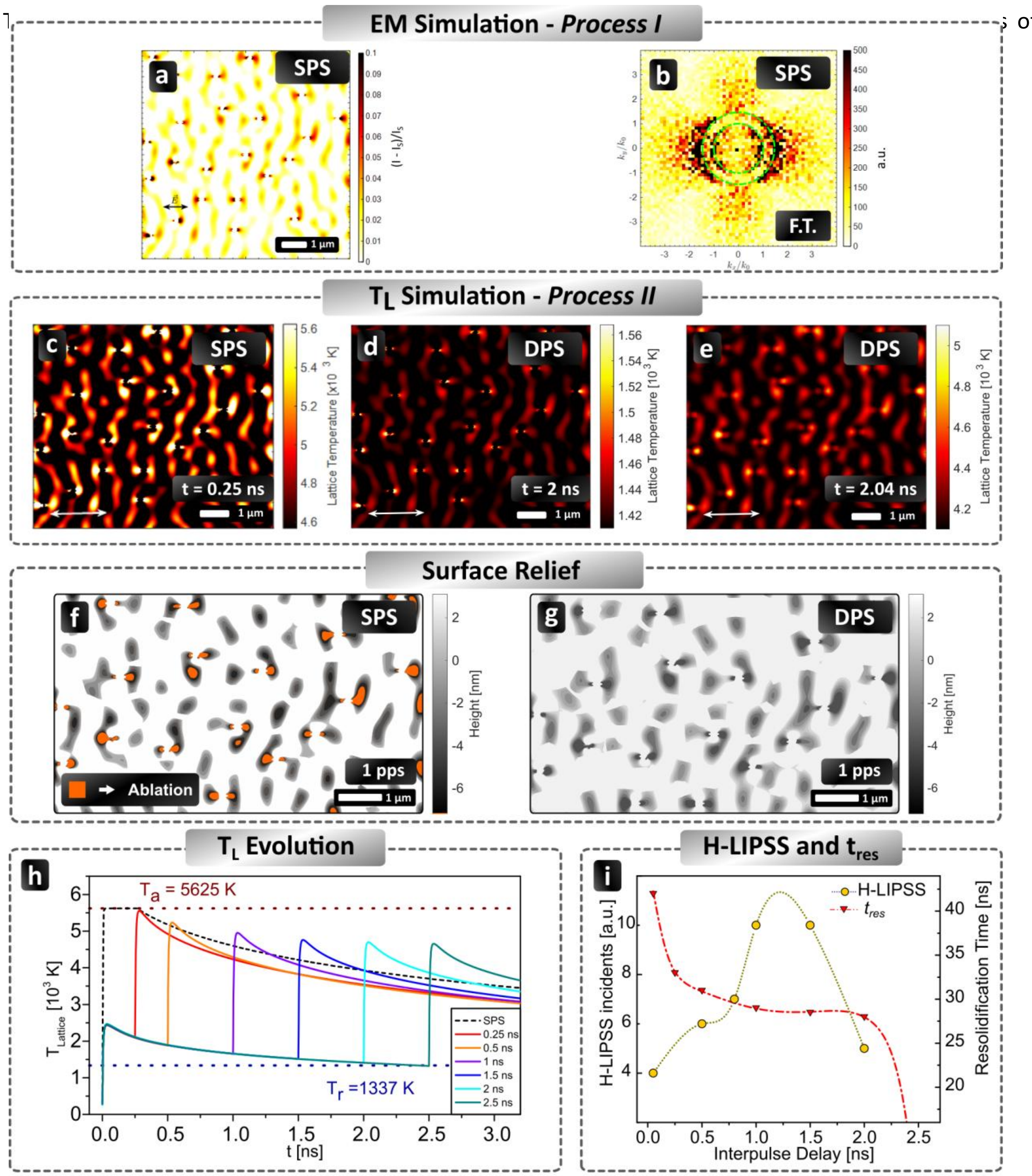

Figure 5: (a) Intensity distribution in the X-Y (transverse) plane just below the top air/Au interface of a 32 nm $Au/SiO_2$ thin film for $\lambda$= 1026 nm, (b) Fourier spectrum of the intensity patterns in the XY-plane showing quasi-periodic features of two distinct periodicities, $\Lambda_{top} \approx \lambda = 1026\, nm$ (inner circle $k_x/k_0 \approx 1$) and $\Lambda_{bottom} \approx 690$ nm (outer circle $k_x/k_0 \approx \lambda/\Lambda$ ). (c) Lattice Temperature spatial profile on the surface of Au for SPS at $t$=0.25 ns (*white* regions show ablated part). Lattice Temperature spatial profile on the surface of Au at $t$=2 ns (d) and $t$~2.04 ns (e) (*white* double-headed arrow indicates the laser polarization). Lastly, simulation results of surface relief resulting from irradiation with $\Phi$ = 280 mJ/cm$^2$ for a single pulse for SPS (f) and a single pair of pulses for DPS (g). *Orange*-colored areas indicate ablation in f. h: Lattice temperature upon irradiation with single and double pulse with variable interpulse delay as indicated. Horizontal lines mark the ablation temperature (*red* dot line, for $T_a$= 0.9*$T_{cr}$= 5625K) and the resolidification temperature (Blue dot line at $T_{melt}$ = 1337K) i: Comparison between resolidification time (simulation data) and number of H-LIPSS within the process window versus the interpulse delay. *Yellow* and *red* dotted lines are added as guyed to eye.

electromagnetic interactions in a spatially incoherent way. In addition, a film damage due to ablation, prevents a homogeneous LIPSS formation. For DPS, the surface is progressively roughened, and the number of defects increases upon increase of the dose, without the formation of craters. In this case, LIPSS are formed prior to the crater formation and, as a result, their spatial coherence is maintained. It is quite likely that the mechanism of LIPSS formation is different between in SPS and DPS cases. In particular, LIPSS in SPS are only formed due to film imperfections, whilst in DPS due to a laser-induced surface roughening effect. A detailed examination of the evolution of Lattice in relation to the $\Delta\tau$ values and the occurrence of ablation and resolidification will be discussed next.

**Temperature evolution of the irradiated solid target**

To analyze the role of the thermal effects in the formation of features of the observed topographies and to elucidate further the underlying processes leading to the formation of H-LIPSS, it is important to conduct a detailed investigation of a thermal response of the irradiated solid target as a function of the pulse separation $\Delta\tau$. In particular, the analysis of evolution of lattice temperature following the laser pulse irradiation can reveal crucial details about the experimentally observed behavior. Theoretical results of the predicted maximum lattice temperatures after the first ($T_{Lmin}$) and the second ($T_{Lmax}$) pulse for a DPS as a function of $\Delta\tau$ are illustrated in Fig.5h.

For DPS and $\Delta\tau$ = 0.250 ns the minimum lattice temperature is $T_{Lmin}$ = 2105 K, whereas the maximum lattice temperature after the second pulse is $T_{Lmax}$ = 5550 K, which is close to the ablation threshold. We recall that the $T_{cr}$=6250 K stands for the critical temperature of Au and $T_a$=0.9$T_c$ (=5625 K) is taken as the minimum lattice temperature above which the material is ablated[46,47]. As the delay increases to $\Delta\tau$ = 0.5 ns, $T_{Lmin}$= 1893 K and $T_{Lmax}$= 4752 K. For $\Delta\tau$ = 1 ns, $T_{Lmin}$= 1656 K which is considerably higher than the melting point $T_{melt}$ (=1337 K)[47] and $T_{Lmax}$ = 4944 K far below ablation temperature. Similarly, for $\Delta\tau$ = 1.5 ns, $T_{Lmin}$= 1500 K and $T_{Lmax}$= 4752 K. Interestingly, for $\Delta\tau$ = 2 ns, $T_{Lmin}$= 1410 K, which is slightly above $T_{melt}$, while for $\Delta\tau$ = 2.5 ns, $T_{Lmin}$= 1320 K that is below the resolidification temperature. This indicates that for $\Delta\tau$ = 2.5 ns, the material has resolidified before the second pulse irradiates the solid.

The above analysis indicates that there are three regimes dictated from the values of the maximum lattice temperatures at various interpulse delays that influence the pattern formation:

(i) when $0.25\ \text{ns} < \Delta\tau < 0.5\ \text{ns}$, $T_{Lmax}$ is close to $T_a$; Ablation may occur and $T_{Lmin}$ is above $T_{melt}$ and therefore the second pulse always irradiates material in the molten phase,

(ii) when $1\ \text{ns} < \Delta\tau < 1.5\ \text{ns}$, $T_{Lmax}$ is well below $T_a$ and $T_{Lmin}$ is above the resolidification temperature and therefore, there is no ablation while the second pulse always irradiates material in the molten phase,

(iii) when $\Delta\tau >= 2$ ns, where $T_{Lmax}$ is well below $T_a$ and $T_{Lmin}$ is very close to $T_{melt}$ and therefore, resolidification may occur before the second pulse arrives.

The impact of $\Delta\tau$ is further elucidated through focusing on the dependence of the resolidification time and the number H-LIPSS occurrences (previously, illustrated in Figure 2) on the interpulse delay (Figure 5i). In particular, in Section IIb, it was shown that the values of the combinations of $\Phi$ and *Ov* leading to H-LIPSS varies significantly and there is a pronounced dependence on the interpulse delay $\Delta\tau$. On the other hand, for $\Delta\tau < 0.5$ ns, a sharp decrease of the $T_r$ from $T_r$ = ~42 ns (SPS) to $T_r$ = ~ 31 ns ($\Delta\tau$ = 0.5 ns). In that $\Delta\tau$ range, the incidents of H-LIPSS process window remains negligibly small (smaller than four H-LIPSS occurences). At higher values of $\Delta\tau$, in the range between 0.5 ns < $\Delta\tau$ < 1 ns the resolidification time drops gradually from $T_r$ = ~31 ns ($\Delta\tau$ = 0.5 ns) to $T_r$ = ~ 29 ns ($\Delta\tau$ = 1 ns) whilst the process window of H-LIPSS grows significantly from four to ten incidents, respectively. Interestingly, the reduction of $T_r$ is very slow for delays in the range between 1 ns < $\Delta\tau$ < 1.5 ns where the H-LIPSS incidents retain the highest value of 10. In this regime, the resolidification time decreases slightly ($T_r$=~28 ns) before a sharp drop when the pulse separation exceeds the single pulse resolidification time (~2.4 ns). A guide-to-eye line is added to visualize the trends in $T_r$ (*red* dotted line) and H-LIPSS cases (*yellow* dotted line). The remarkable delay in the resolidification process compared to bulk materials (for bulk materials, resolidification is expected within a few ns) is due to the fact that the small thickness inhibits the electron diffusion which, in turn, is reflected on the ultrafast dynamics and thermal response of the material.

To summarise, the above analysis reveals that the H-LIPSS process window maximization occurs for 1 ns < Δτ < 1.5 ns; In this range, the resolidification time is almost constant. Furthermore, $T_{Lmax}$ is well below $T_a$ and $T_{Lmin}$ is above the resolidification temperature and therefore, there is no ablation while the second pulse always irradiates a material in a molten phase.

### *f) Optical properties of laser-structured films and biosensing prospects*

The fabrication of relatively homogeneous periodic arrays in a large area gives a promise for the excitation of ultranarrow Surface Lattice Resonances, which arise due to diffraction coupling of individual localized plasmon resonances over nanoparticles/nanostripes[8], and are extremely promising for biosensing and other applications[8]. Here, the larger is a quasi-homogeneous regular area, the narrower are the resonances. Typically, the diffraction coupling of 100 and more resonators is sufficient for obtaining relatively narrow SLRs, while the coupling of 1000 and more resonators can reduce the resonance width down to 1-2 nm FWHM[8]. In our study, we examined optical properties of laser-structured LIPSS in order to check whether the structural quality and regularity of formed arrays are sufficient to excite SLRs with high resonance quality. We recorded a pair of ellipsometric parameters, $\Psi$ and $\Delta$, in the wavelength range from 250 nm to 1700 nm as a function of angle of incidence as described in the Methods section.

In our tests, we used an optimized structure produced by DPS at fluence of $\Phi$ = 170 mJ/cm$^2$ and time delay Δτ = 500 ps (Fig. 1). As follows from a magnified SEM image of this sample (Fig. 6a), the formed LIPSS presented a regular array of nanostripes with the period of Λ = 655 ± 20 nm. As one can see from an AFM image (Fig. 6b), the depth of trenches between nanostripes was about 30 nm, which indicates a drastic surface reorganization, potentially leading to a complete removal of the 32-nm gold layer under laser structuring. Fig. 6c and Fig. 6d show the ellipsometry parameters measured on this sample. Here, one can identify three sets of collective resonances observed at wavelengths of ~700 nm, ~1250 nm and ~1600 nm. While relatively weak resonances around 700 nm and 1600 nm can be attributed to coupling of incident light to running plasmons over gold, a pronounced resonance around 1250 nm is obviously due to diffraction coupling of localized plasmons excited in the fabricated periodic nanostructure. Indeed, these essentially plasmonic SLRs have a spectral position described by the formula $\lambda_{\mathrm{Res}} = a\left(n + \sin\theta\right)$, where *a* is the period of the structure, *n* is the effective refractive index for the light beam diffracted along the surface of the sample and $\theta$ is angle of incidence[8]. As shown in Fig. 6e, this formula provides a nice fit to the measured resonance position if the parameters are selected as follows: *a* = 650 nm and *n* = 1.06. In this case, the period of the structure calculated from optical measurements is in an excellent agreement with the period extracted from SEM images, while the refractive index is close to the refractive index of air (which suggests that the diffracted beam is only slightly affected by the presence of underlying gold LIPSS). It is worth noting that the optical properties of the examined sample also cannot be described by simple Fresnel theory due to the presence of diffracted waves[8]. One can also see from Fig. 6c that SLR features are really narrow with the resonance spectral width of about 20 nm full width half maximum (FWHM), corresponding to the resonance quality factor of Q = 65, which is much higher than the relevant value in the case of uncoupled LPR[3] (the resonance width for LPRs is about 80-100 nm FWHM, corresponding to Q ~ 10). The quality factor value Q = 65 is consistent with the presence of about 200 nanostrip-based resonators within the illumination spot of the used ellipsometer system (650 nm LIPSS period over a 30*60 $\mu m^2$ focal spot). It should be noted that the spectral sensitivity of SLRs to refractive index variations is related to the diffractive nature of light coupling to plasmons[8], linking the sensitivity to the structure periodicity as $\Delta\lambda/\Delta n \sim a$. This means that the sensitivity in our case should be around 650-750 nm of spectral shift to RIU variations. Nevertheless, a small width of SLRs should improve significantly the performance of plasmonic biosensors. To take into account the sharpness of the resonance and thus examine the ability of the system to sensitively measure small wavelength changes, one normally uses a characteristic "Figure of Merit" (FOM) parameter FOM=$(\Delta\lambda/\Delta n)\cdot(1/\Delta\omega)$, where $\Delta\omega$ is the width of resonance at FWHM and $\Delta\lambda$ is the resonance shift for a $\Delta n$ refractive-index change[48]. The essence of FOM consists in an adequate quantification of sensing potential of a sensing transducer in configurations similar to those used in commercial instruments. Typical FOMs do not exceed 8 in the case of uncoupled LPR[3] and ~20 in the case of SPR[49]. As we showed in a previous study, the employment of SLRs of similar resonance width enables one to increase these parameters up to 120 and more[50]. The employment of so narrow SLRs in regular metamaterial arrays can enable at least one order of magnitude better sensing performance compared to uncoupled LPRs in disordered particle arrays[51], which highlights the importance of observed SLRs for biosensing applications.

Another important observation is related to the presence of jumps of light phase at the minima of SLRs. As shown in Fig. 6d, the jumps were generated in the very minima of SLRs, while the total amplitude of such phase jumps was not very large (less than 8° Deg. of phase). It should be noted that phase jumps appear under any sudden drop of light intensity in reflection (light darkness)[6,52,53]. Here, the lower the intensity at the resonance minimum, the sharper the phase jump[6]. In our case, the resonances were not very deep, which conditioned relatively small values of phase jumps. Such a limitation in terms of the depth of observed resonances is straightforward. As we showed in previous studies[5,6,9] zero intensity (complete light darkness) in regular nanoparticles/nanostripes arrays can normally be achieved under relatively large size of plasmonic features (typically, 100-130 nm). In our case, the size of laser-structured features was comparable with the original thickness of the Au film (~ 32 nm), which conditioned essentially non-zero light intensities at the resonances and related modest values of phase jumps (although the resonances were very narrow due to diffraction-coupling effect). We believe that further optimizations of the technique of femtosecond laser structuring would allow one to structure thicker films in a similar manner in order to form regular arrays composed of larger plasmonic elements (as the optimum, 100-120 nm). This will make possible the combination of ultra-narrow SLRs and the generation of phase singularities. In addition, for such arrays we expect the generation of topologically dark quasi-resonant features, not related to SLRs, which make possible a much enhanced (virtually unlimited) spectral sensitivity to RI variations due to "scissor effect" described in our recent study[54]. Based on already obtained experience and results, the fabrication of such structures looks feasible, although it will require a further elaboration of the process. This work is now under progress and will be published elsewhere.

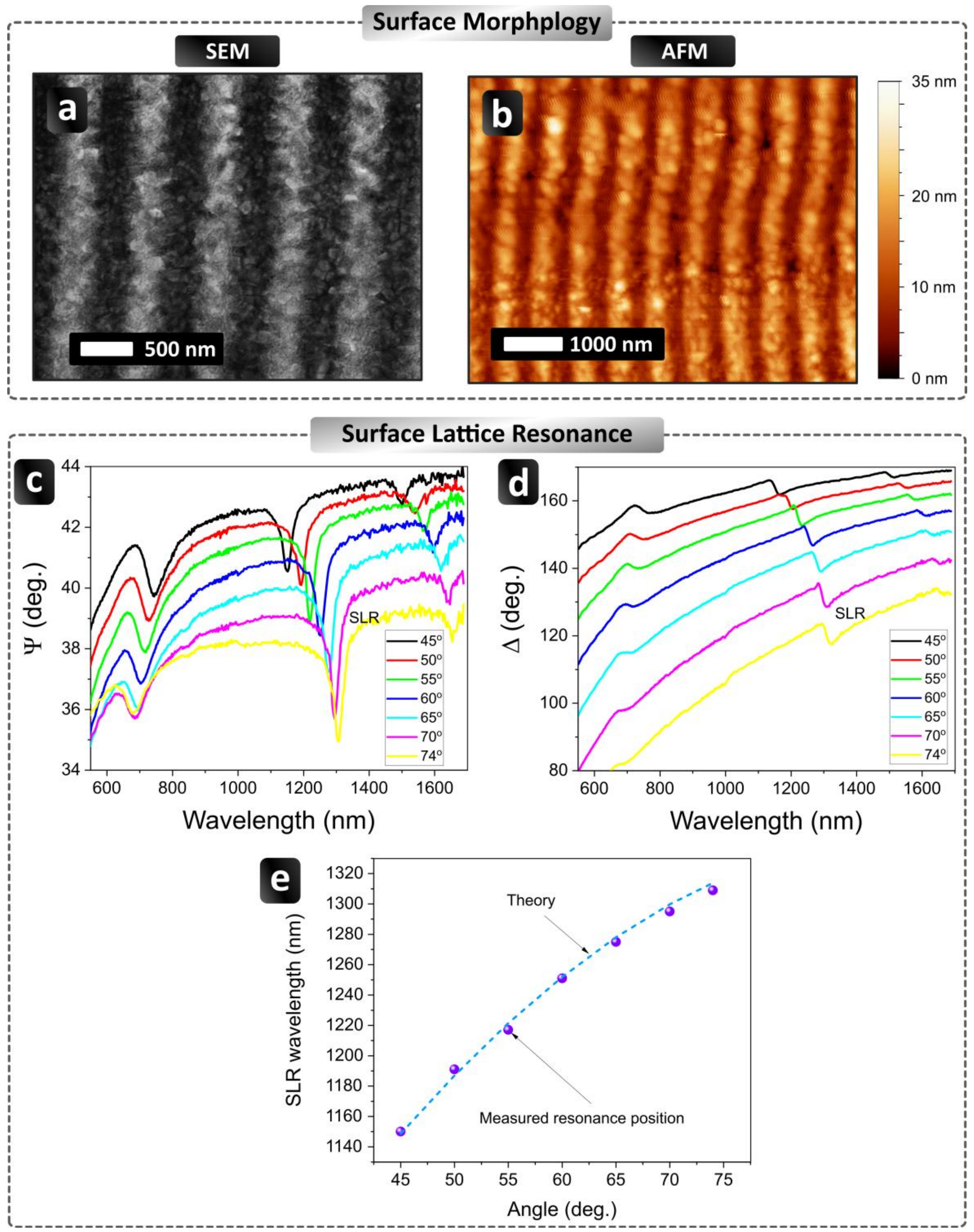

Figure 6. SEM (a) and AFM (b) images of a sample prepared at fluence of $\Phi$ = 170 mJ/cm$^2$ and time delay $\Delta\tau$ = 500 ps. (b) Ellipsometric parameter $\Psi$ (amplitude) as a function of the incident wavelength and angle for the sample shown in panels (a) and (b). SLR denotes the positions of the plasmonic surface lattice resonances. (c) Ellipsometric parameter $\Delta$ (phase) as a function of the incident wavelength and angle for the same sample. (d) Comparison of the measured resonance position of SLR (magenta circles) and theory (dotted line).

Thus, we reproduced the generation of narrow diffraction-coupled SLRs and related phase features using regular periodic nanostripe-based structures fabricated by a technique of femtosecond laser processing, which is much less costly than EBL and FIB technologies in the fabrication of large area arrays and scalable. Indeed, laser processing does not require highly costly high vacuum and lithography steps and can be easily programmed for prompt structuring of a gold film over a large area. It should be mentioned that different laser-assisted methods were previously used for the fabrication of plasmonic arrays, making possible the excitation of plasmonic resonances and related abrupt phase features. However, all previous approaches typically used extra photo- or nanoparticle lithography steps,[55,56] or electroless plating step in the case of 3D plasmonic arrays[57] which complicates the whole fabrication procedure. In this study, we avoided lithography steps and demonstrated the fabrication of such plasmonic arrays by a direct laser structuring of a thin metal film, which opens up avenues for a large-scale manufacturing of phase-sensitive plasmonic transducers for ultrasensitive biosensing.

# III. Conclusions

The results derived in this work demonstrate the possibility of generating plasmonic sensing elements on thin Au film upon femtosecond laser structuring of a glass substrate-supported thin Au film. To this end, a systematic study of LIPSS formation on thin Au film upon single and double pulse ablation is presented. The formation of highly regular LIPSS over large areas on thin Au film surface upon double pulse structuring is investigated, and the underlying formation mechanism is discussed. In particular, the key role of the interpulse delay in producing regular LIPSS structures over large areas is discussed and the optimum interpulse delay regime is identified to range between $\Delta\tau$ = 1 ns and $\Delta\tau$ = 1.5 ns. The striking differences in the outcome of single and double pulse processing underline the hydrodynamic origin behind the regularity of LIPSS formation. Electromagnetic simulations of the propagation of the incident laser beam on Au surface provides insight on the origin of LIPSS periods and theoretical predictions appear to agree with the experimental data. Furthermore, simulations of the lattice temperature evolution elucidate the essential impact of DPS of a particular delay regime, in delivering sufficient energy to texture the material without initiating ablation. The potential role of convection flow in homogeneous LIPSS fabrication was underlined. Furthermore, ellipsometry measurements validate the possibility of using the formed LIPSS architectures as transducers in ultra-sensitive plasmonic biosensing. Indeed, the reflection by the regular LIPSS areas exhibits surface lattice resonances having a very narrow spectral width (~20 nm) and related phase features at resonance minima, which look very promising for biosensing and other applications. We believe that our results provide comprehensive and valuable data for the generation of functional periodic structures on Au thin layers and can further contribute to the development of new applications of laser functionalized surfaces.

# IV. Materials and methods

## *a) Laser processing*

Thin Au films with thickness of $d_{Au}$=32 ± 2 on glass substrate ($d_{Sub}$=170 µm) have been laser processed using the radiation of a 170 fs laser source emitting at 1030 nm having a repetition rate of 5 kHz (Light

Conversion, Pharos). For the generation of the double pulses and the tuning of a time delay $\Delta\tau$, a setup shown in Figure 7 was developed. In detail, the main beam is divided into two parts by a polarizing beam splitter (BS) and then is guided into two arms, Arm A and Arm B respectively. Arm A is controlled by a computer-assisted micrometer displacement controller (DC) used for setting the $\Delta\tau$ value in the range of 0 to 50 ps with accuracy of 2 fs. Arm B is placed on a manual dovetail rail to produce $\Delta\tau$ values in the range of 100 ps to 2 ns with an accuracy of ~3.5 ps.

The first half wave plate (λ/2) combined with a linear polarizing cube (LPC) is utilized to control the fluence distribution between the two pulses. The second λ/2, in Arm B, controls the polarization of Arm B, which in this series of experiments is fixed vertically as shown in Figure 7. As a result, the two arms generate two pulses with parallel polarization incident onto the sample surface. The recombination of the beams of the two arms in colinear propagation is realized by the BS. Beam attenuation is realized via a computer controlled λ/2 in combination with a LPC. A computer-controlled attenuation part consisting of a half waveplate and a LPC is used to tune the pulse energy. A polarization control unit consisting of a λ/2 is used to control the polarization direction in the sample plane. The sample is placed on a programable 3-axis motorized stage. The beam was focused on the sample by a f = 100 mm plano convex lens giving a spot size $2w_0$ of ~55 μm in diameter at $1/e^2$ using a CCD camera placed at the focal plane. The overlap of the two pulses is estimated with accuracy in the order of the pulse duration (170 fs), utilizing a second harmonic generation (SHG) crystal. The experiments were conducted at normal incidence and in ambient air. The process parameters considered in the experimental part are: The peak fluence, $\Phi$ that is calculated using the equation (1) in Table 1 where the $E_p$ is the energy per pulse measured with a pyroelectric power meter. Fluence $\Phi$ value refers to the total fluence of a pulse set in case of DSP and the fluence of a single pulse in case of SPS. The overlap (Ov) is estimated as the average number of incident pulses at any point of irradiated area (pps) as describe in the equation (2) of Table 1, where $u$ is the average speed of the moving stage and $f$ the repetition rate. Finally, $\Delta\tau$ is calculated using equation (3) Table 1, where $c$ stands for the speed of light, $\Delta L$ is the optical path difference of the two beams while the presence of the factor of 2 comes from the fact that the pulse travels two times the displacement of the arm due to the setup geometry. The hatch, $H$ is defined as the step between the scanning lines and is fixed to $H$ = 2 μm, after some preliminary study, unless otherwise stated. Finally, the average dose $D = \Phi \cdot pps$ is a measure of the total irradiation energy per spot.

Table 1:1 Process parameters and value ranges

| Value | Symbol | Equation | Min | Max |
|---|---|---|---|---|
| Peak Fluence | $\Phi$ | (1) $\Phi[J/cm^2] = 2E_p/\pi w_0^2$ | 80 mJ/cm² | 320 mJ/cm² |
| Pulse Overlap | ***Ov*** | (2) $Ov[pps] = (2w_0/u) \cdot f$ | 20 pps | 400 pps |
| Hatch | ***H*** | | 2 μm | 20 μm |
| Interpulse delay | $\Delta\tau$ | (3) $\Delta\tau\ [ps] = 2 \cdot \frac{\Delta L[\mu m]}{c} \cdot 10^6$ | 0 ps, 5 ps, 10 ps, 20 ps, 50 ps, 100 ps, 300 ps, 500 ps, 800 ps, 1 ns, 1.5 ns, 2 ns. | |

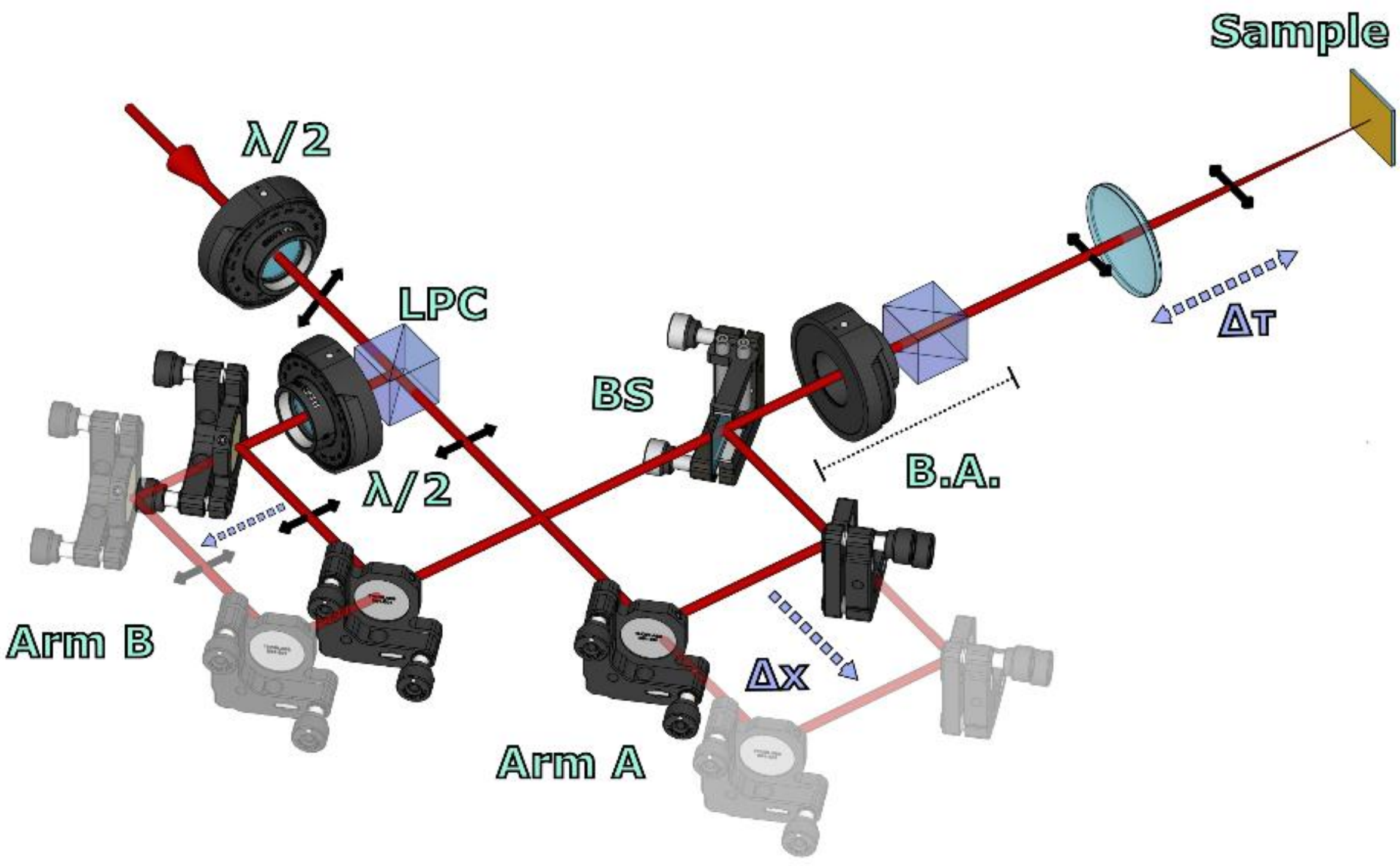

Figure 7. The setup. Abbreviations: beam splitter (BS), linear polarizer (LPC), half waveplate (λ/2), Attenuation part (AT), polarization control part (PC). Linear displacement (Δx), time delay ($\Delta\tau$).

The morphologies of the laser-fabricated structures were visualized by a field-emission Scanning Electron Microscope (SEM). All the measurements of the features of surface structures were performed by a 2D-FFT analysis of the corresponding SEM images using Gwyddion (http://gwyddion.net/), a free and open-source software for data visualization and analysis.

### *b) Electromagnetic Simulations*

Here we provide a description of the theoretical model used for the determination of the electromagnetic modes that are excited after irradiation of a thin Au film with fs pulses. Analytical approaches such as the Sipe theory, provides a *near-surface* description of the inhomogeneous absorption of optical radiation by the roughened surface[58]. On the other hand, computational approaches based on the numerical solution of the Maxwell equations such as finite-difference time-domain (FDTD)[59] and finite integration technique (FIT)[60] are capable to provide details about the electromagnetic modes produced on complex media in three dimensions. We investigate numerically the spatial modulation of the energy below the irradiated rough surface of a $d = 32$ nm thick Au/$SiO_2$ thin film, resulting from the electromagnetic field distribution i.e. SPP coupling between the two dielectric-metal interfaces and other surface waves, by solving the integral form of the Maxwell equations. For this purpose, the Maxwell's Grid Equations are solved considering a three-layer system (air-metal-glass) with optical constants $\varepsilon_a = 1$ (air), $\varepsilon_g = 2.1$ (glass) and $\varepsilon_m = -44.47 + i3.2$ (Au)[61] for laser wavelength $\lambda_L = 1026$ nm. For Au thin film described above, the optical skin depth is comparable to the film thickness. The laser beam is considered to be a normally-incident plane-wave and linearly polarized along the $x$−axis of duration $t$ =100fs. The laser beam is propagating along the $z$ axis and the $xy$ plane, i.e. the sample plane, is perpendicular to the propagation direction. We keep simple periodic boundary conditions for $xy$, while at the $z$-boundaries normal to the propagation, convolutional perfectly matched layers (CPML) are used in order to truncate the computational domain and avoid non-physical reflections at the edges of the simulation grid. The periodic effects due to periodic boundary conditions are of low importance since the irradiated metallic area is large enough and can be supressed by the optical losses. Using the boundary conditions described above, the spatial laser profile matches perfectly the $xy$ plane of the structure with homogeneous intensity distribution. The surface roughness plays the key role in electromagnetic wave scattering and generation of SPP that are the precursors of LIPSS. To emulate the features of a rough surface, we introduce randomly distributed scattering centres in the form of nano-holes of radius $r < d$ along the film surface. The concentration of inhomogeneities at the air/Au interface is considered to be $C = N\pi r^2/S \approx 0.25\%$ where N is the number of the nano-holes and the laser affected area $S = 10 \times 10$ μm$^2$. A similar approach has been introduced in similar studies for bulk materials[34,62,63]. The interaction of light with the surface inhomogeneities of the material produces electromagnetic interference patterns along the surface, which determine the energy absorption landscape. Since the absorbed energy of the film is proportional to the intensity, in order to capture the absorption energy maxima and minima due to scattering of overlapping surface waves, we calculate the normalized intensity difference $(I - I_S)/I_S$ where $I \sim |\vec{E}|^2$ is total intensity of Au films below the rough surface, while $I_S$ is the total intensity of Au below a defect-free surface. This difference depicts the intensity maxima and minima due to both scattered radiative and non-radiative fields by the subwavelength imperfections.

### *c) Ellipsometric Measurements*

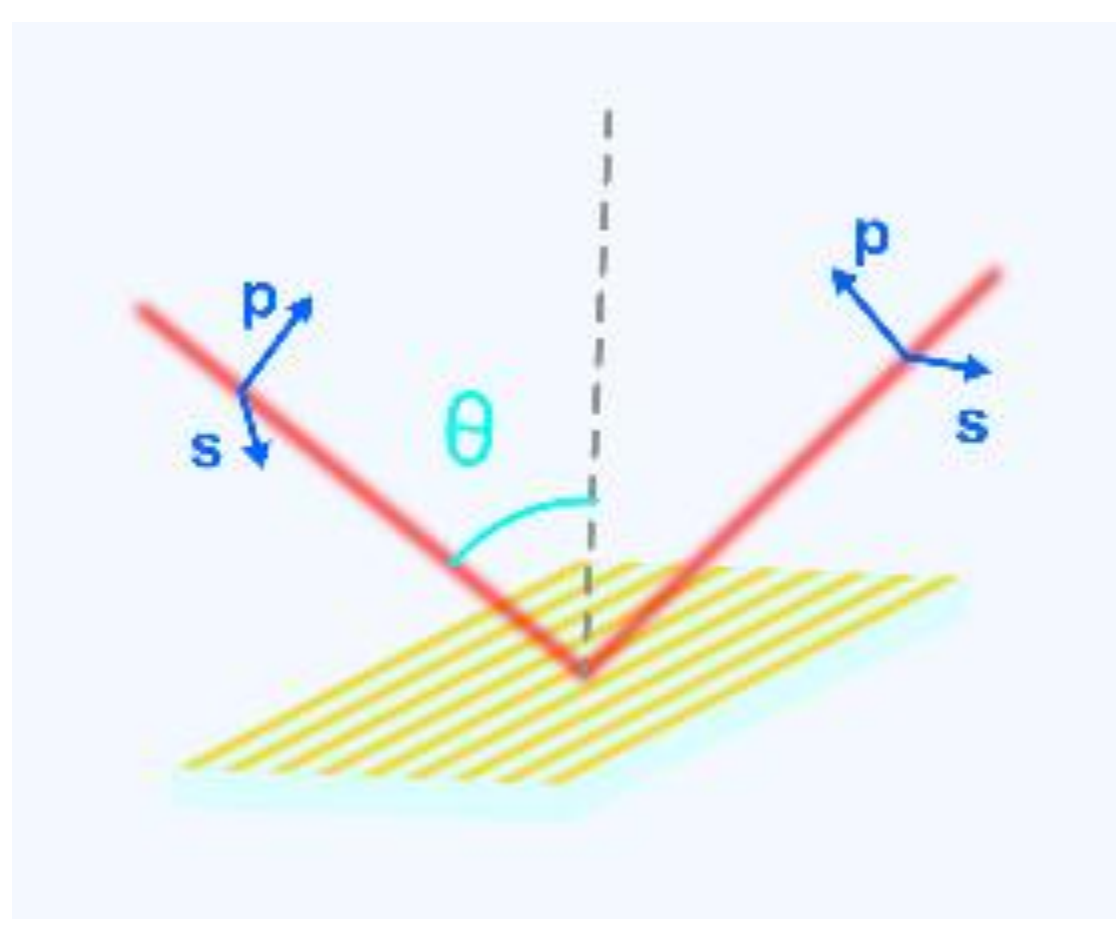

Figure 8 Schematic diagram of ellipsometric measurement for gold patterned nanostructures and geometry of orientation of ablated samples with respect to the direction of incident polarised light.

To measure optical properties of the fabricated samples, we used a variable angle focused-beam spectroscopic ellipsometer Woollam M-2000F. It is based on the rotating polariser-compensator-analyser setup (Figure 2) and utilises a diode array spectrophotometer to extract the spectral parameters Ψ (ellipsometric reflection) and Δ (ellipsometric phase) in the wavelength range of 240-1690 nm, with a wavelength step of ~1.0 nm for 240-1000 nm and ~2.0 nm for 1000-1690 nm. The beam spot size on the sample was approximately 30 μm × 60 μm for ~60-70° angles of incidence. These parameters are related to the sample reflection as $\tan(\Psi)\exp(i\Delta) = r_p/r_s$, where $r_p$ and $r_s$ are the amplitude reflection coefficients for *p*- and *s*-polarized light, respectively. In addition to ellipsometric parameters Ψ and Δ, the ellipsometer enables one to separately measure $R_p=|r_p|^2$ and $R_s=|r_s|^2$ providing the intensity reflection spectra for *p*- and *s*-polarised light, respectively. The errors on the polariser, compensator and analyser azimuthal parameters are generally less than ±0.01°, which implies that, the ellipsometric parameters Ψ and Δ can be measured with an error of level ±0.02°. The schematic for ellipsometry measurements are presented in 2, showing that the Au plasmonic LIPSS were oriented in such a way that the plane of incidence was parallel to the array lattice vector (i.e the LIPSS were perpendicular to the plane of incidence). An Ultra Plus Carl ZEISS SEM was used for high-resolution imaging of the nanostructures. The unique in-lens SEM detector gives resolution of the order of 1.0 nm at 15 kV (1.6 nm at 1 kV), dependent on the type of samples.

Theoretical modelling of optical properties of fabricated samples was performed with the help of Fresnel theory where we applied the effective-medium approximation (EMA) to the top metal layer nanostructured by laser ablation. WVASE32 software of J.A. Woollam Company was used to perform calculations. The model geometry for studied samples was chosen to be constructed of three layers: glass as a substrate, unperturbed film of Au and an EMA layer of ablated gold film which was a combination of gold and air. The optical constants of EMA layer were calculated using a Maxwell-Garnett theory. The thickness of the unperturbed gold layer, thickness of the EMA layer and the ratio of void to gold in the EMA layer were varied to achieve the best fit with the measured data.

## Acknowledgements

This work was supported by the EU's H2020 framework programme for research and innovation under the NFFA-Europe-Pilot project (Grant No. 101007417). VGK and ANG acknowledge support of Graphene Flagship programme, Core 3 (881603). AVK, FF and ES acknowledges contribution of the International Associated Laboratory (LIA) “Laser Matter Interaction from fundamental studies to inNOvative laSer processing- MINOS”. AVK acknowledges support from the French government under the France 2030 investment plan, as part of the Initiative d’Excellence d’Aix-Marseille Université – A*MIDEX ” AMX-22-RE-AB-107.